\documentclass[sigconf]{acmart}

\AtBeginDocument{%
  }

\usepackage{tcolorbox}

\usepackage{bbding}
\usepackage{subfigure}
\usepackage[boxed,ruled,lined]{algorithm2e}
\usepackage{multirow}
\usepackage{makecell}

\usepackage{arydshln} 
\usepackage{enumerate}
\usepackage{enumitem}
\usepackage{makecell}
\usepackage{bbm}

\newcommand{\ourname}{SSRLive\xspace}

\newcommand{\paratitle}[1]{\vspace{0.8ex}\noindent\textbf{#1}}

\setcopyright{acmlicensed}
\copyrightyear{2018}
\acmYear{2018}
\acmDOI{XXXXXXX.XXXXXXX}
\acmConference[Conference acronym 'XX]{Make sure to enter the correct
  conference title from your rights confirmation email}{June 03--05,
  2018}{Woodstock, NY}
\acmISBN{978-1-4503-XXXX-X/2018/06}

\begin{document}

\title{\ourname: Live Streaming Recommendation with \\
Dynamic Semantic ID}

\author{Teng Shi$^{1}$, Zhaoheng Li$^{1}$, Yuanhang Qu$^{1}$, Yi Liu$^{1}$, Lixiang Lai$^{1}$, Yuning Jiang$^{1}$}
\email{{yunshi.ly, lixiang.llx, mengzhu.jyn}@alibaba-inc.com}
\affiliation{
\institution{$^{1}$Taobao \& Tmall Group of Alibaba, Beijing, China}
  \city{}
  \country{}
}

\renewcommand{\shortauthors}{Shi et al.}

\begin{abstract}
Live streaming has emerged as one of the fastest-growing forms of online media, enabling instant content broadcasting and real-time engagement between users and streamers. 
Despite the effectiveness of existing recommendation algorithms in this domain, they often suffer from limited utilization of computational resources, with low FLOPs that hinder further performance enhancement. 
Generative recommendation techniques, which have gained traction in various industrial tasks, offer a promising avenue for improving live streaming recommendations. 
However, directly applying generative methods to live streaming is non-trivial due to two major challenges: 
(1)~static semantic IDs (SIDs) cannot reflect the rapidly changing nature of live room content; and 
(2)~generative pipelines generally do not incorporate user--streamer interaction signals (e.g., likes, orders), which are critical for modeling user intent toward both the streamer and showcased products. 
To address these challenges, we introduce \textbf{\ourname}: Dynamic \textbf{S}emantic ID-guided \textbf{S}treaming \textbf{R}ecommendation for \textbf{Live} platforms. 
The proposed framework integrates a generative module and a discriminative module in a unified architecture. 
The generative component employs an encoder--decoder design to produce both static and dynamic SIDs, enabling timely representation of live room content while leveraging multimodal information. 
The discriminative component refines task-specific representations by combining SIDs with user features, augments them with user--streamer interaction data, and performs multi-task predictions. 
Extensive offline experiments on production-scale datasets show that \ourname consistently surpasses competitive baselines, while online A/B tests in real-world deployment demonstrate tangible benefits: watch time (+3.38\%), GMV (+0.72\%), follower growth (+3.12\%), and interaction volume (+2.92\%). 
These improvements highlight the effectiveness and business value of \ourname, which is now fully deployed, serving hundreds of millions of active users.

\end{abstract}



\maketitle

\section{Introduction}
With the rapid growth of mobile Internet technologies, live streaming has become a key form of digital media, enabling real-time broadcasting by streamers and user-streamer interaction. Alongside short videos and digital news, it holds rising societal and commercial importance, driving research interest in live streaming~recommendation~\citep{lu2025liveforesighter,liu2025llm,deng2024multimodal,cao2024moment}.

Recent studies have explored various aspects of live streaming recommendation,  
including capturing interaction patterns between users and streamers~\citep{zhu2025live},  
representing the multimodal content of live rooms~\citep{deng2024multimodal},  
and modeling dynamic contextual factors inherent to live streaming.  
Representative examples include highlight detection~\citep{zhao2022antpivot} and forecasting future live streaming content~\citep{lu2025liveforesighter}.

Despite the promising results of existing live streaming recommendation algorithms, several challenges persist. 
(1)~Traditional Deep Learning Recommendation Models (DLRMs), typically built on attention mechanisms and multilayer perceptrons (MLPs)~\citep{DIN,naumov2019deep,chang2023pepnet}, already deliver strong online performance, and further architectural refinements often yield only marginal gains. 
(2)~Current DLRMs do not efficiently utilize computing resources~\citep{OneRecV1}, exhibiting low FLOPs and limited Model FLOPs Utilization (MFU) in practice. 
These limitations hinder further performance improvement in live streaming recommender systems.

In recent years, large language models (LLMs) have shown exceptional performance across diverse and complex NLP tasks~\citep{zhao2023survey}. 
Motivated by these advances, researchers have adapted core LLM techniques to recommender systems, driving the emergence of generative recommendation~\citep{TIGER,LC_Rec}. 
Many approaches have been deployed in large-scale industrial settings with promising results~\citep{OneRecV1,HSTU,GPR,MTGR,Rankmixer}. 
These successes stem primarily from:
(1)~Scalability: Transformer-based models can improve online metrics simply by increasing the number of model parameters~\citep{OneRecV1,Rankmixer}.
(2)~Semantic IDs (SIDs): Representing items with semantic IDs~\citep{Forge,OneRecV1} enables integration of multimodal signals and captures item–item correlations more effectively than conventional item IDs.

\begin{figure}[t]
\centering
\includegraphics[width=0.85\columnwidth]{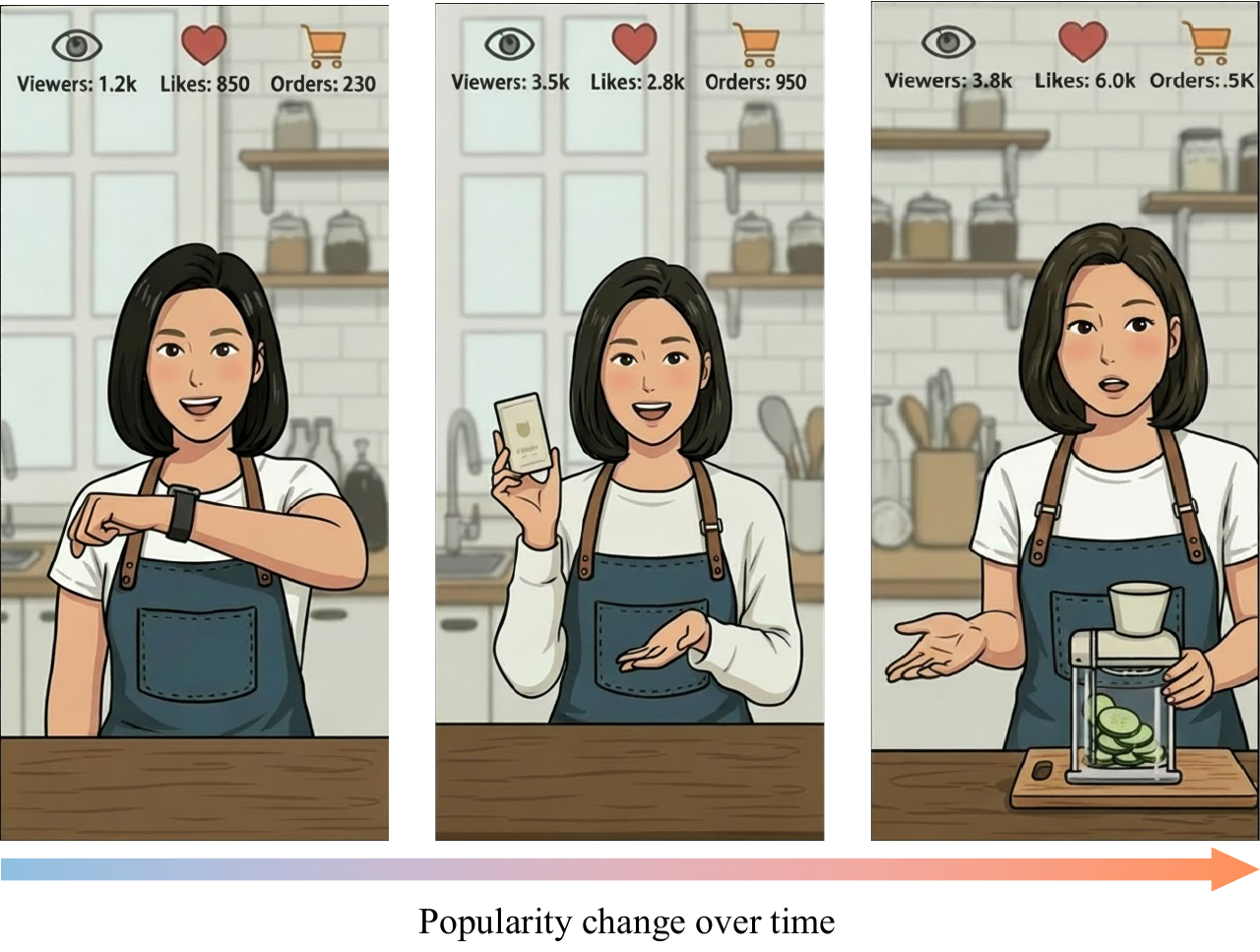}
\vspace{-12px}
\caption{
A live streaming example where a streamer is presenting products. 
The scenario highlights two key characteristics of live streams: 
(1) \textbf{Real-Time Dynamics}: where the streaming content and the popularity vary over time.
(2) \textbf{User–Streamer Interactions}, such as users liking the stream and purchasing products being introduced by the streamer, which reflect users’ preferences toward the streamer.
}
\label{fig:intro_example}
\vspace{-0.5cm}
\end{figure}

The demonstrated success of generative recommendation methods provides a promising direction for advancing live streaming recommender systems. 
However, applying such methods to live streaming scenarios faces several key challenges.
(1)~\textbf{Static SIDs are misaligned with the real-time nature of live streaming.} 
Existing SIDs, such as those applied to short video scenarios~\citep{OneRecV1}, are static and invariant, whereas live streaming content changes in real time. Static SIDs cannot adequately capture the real-time information of a live room.
(2)~\textbf{Purely generative approaches often fail to explicitly model user--streamer interactions.} 
Interactions between users and streamers, such as likes and orders, are vital for capturing user preferences, especially in e-commerce scenarios. 
However, many existing generative frameworks fail to effectively incorporate the rich information conveyed by these cross-interaction signals~\citep{OneRecV1,Forge}.
Figure~\ref{fig:intro_example} provides a concrete example that demonstrates the real-time characteristics of live rooms and underscores the significance of user–streamer interactions.

To address the aforementioned challenges, we propose \textbf{\ourname}:  
Dynamic \textbf{S}emantic ID-guided \textbf{S}treaming \textbf{R}ecommendation in \textbf{Live} Scenarios. 
\ourname adopts a hybrid generative–discriminative architecture to jointly capture the real-time variations in live streaming content and the interaction information between users and streamers.
(1)~\textbf{Generative module.} 
We construct both static and dynamic SIDs. 
The static SID is derived from the streamer's static multimodal information, while the dynamic SID reflects real-time changes in live room content, enabling the model to more effectively represent the temporal dynamics inherent in live streaming scenarios. 
An encoder–decoder architecture is employed to generate both the static and dynamic SIDs.
(2)~\textbf{Discriminative module.} 
To effectively model interaction features between users and streamers, unlike existing approaches~\citep{OneRecV1,Forge} that directly generate SIDs for retrieval, we leverage the SID as auxiliary information to strengthen downstream discriminative tasks. 
For each task, multiple learnable queries are utilized to extract task-specific representations from the SIDs and user features. 
The extracted representations are then fused with live room features and cross features, and the resulting fused representation is fed into multi-task prediction layers to get the final~outputs.
Our main contributions are as follows:

\noindent\textbf{$\bullet$}~We introduce \ourname, a hybrid generative–discriminative architecture.  
The generative module generates static and dynamic SIDs to capture real-time characteristics of live streaming,  
while the discriminative module extracts task-specific information from SIDs and integrates it with cross features to model user–live interactions.

\noindent\textbf{$\bullet$}~Extensive offline experiments and analytical studies on industrial datasets show that \ourname outperforms both our online baseline and representative recommendation models.

\noindent\textbf{$\bullet$}~Online A/B testing in real-world live streaming scenarios demonstrates substantial gains:  
+3.38\% in user watch time, +0.72\% in gross merchandise value (GMV), 
and increases in interaction metrics, including +3.12\% follower growth and +2.92\% active interactions, highlighting its practical impact.
\section{Related Work}
\label{sec:Related_Work}

\paratitle{Generative Recommendation.}
The rise of large language models (LLMs)~\citep{zhao2023survey,Gpt4,DeepseekR1} has driven advances in generative recommendation, applied to retrieval, pre-ranking, and ranking stages with promising industrial results. 
Most work focuses on the \textit{retrieval} stage~\citep{COBRA,Tbgrecall}, where items are encoded as semantic IDs (SIDs) and retrieval is reformulated as a next-token prediction (NTP) task, generating candidate SIDs~\citep{TIGER,LC_Rec,Forge}. 
Others improve retrieval via multi-step reasoning, either explicitly (e.g., OneRec-Think~\citep{OneRecThink}) or implicitly (e.g., ReaRec~\citep{ReaRec}, OnePiece~\citep{OnePiece}). 
In the \textit{ranking} stage~\citep{OneTrans,GenRank}, LLM-inspired methods boost performance by scaling parameters and enhancing cross-feature interaction, as in MTGR~\citep{MTGR} and RankMixer~\citep{Rankmixer}. 
Recently, \textit{end-to-end} generative recommendation~\citep{EGA,OneRecV0} has emerged. 
Unlike traditional multi-stage cascades constrained by earlier-stage outputs, these models jointly optimize the entire pipeline by combining large-scale generative modeling with reward signals, exemplified by systems such as OneRec~\citep{OneRecV1,OneRecV2}, OneSearch~\citep{OneSearch}, and GPR~\citep{GPR}.
Unlike existing static SID methods, we employ dynamic SIDs to model real-time information in live streaming.

\paratitle{Live Streaming Recommendation.}
Live streaming integrates real-time content delivery with rich user–streamer interactions~\citep{cao2024moment,deng2024multimodal,qu2025bridging}, posing challenges distinct from short video recommendation~\citep{li2025farm,liang2024ensure,liu2025llm,lu2025liveforesighter}. 
Key characteristics include: 
(1)~timeliness: live rooms exist only during broadcasts;  
(2)~dynamics: content and user preferences evolve through real-time interactions; and  
(3)~multi-entity relationships: involving users, streamers, live rooms, and often products in e-commerce~\citep{jin2022watch,yu2021leveraging}, limiting the effectiveness of static feedback approaches.
Research spans streaming data modeling and context adaptation.  
Data modeling addresses 
repeated behaviors, 
multimodal fusion of video, audio, and text~\citep{deng2023contentctr,deng2024multimodal,xi2023multimodal,xu2024memf}, and cross domain transfer from short video signals to mitigate sparsity problem~\citep{cao2024moment,zhang2023cross,zheng2023dual}.
Context adaptation captures temporal and contextual changes through highlight detection~\citep{deng2024multimodal,zhao2022antpivot}, and forecasting streamer actions~\citep{lai2023learning,lu2025liveforesighter,zhang2021deep,zhang2022predicting}. 
Recent trends focus on real-time, 
multimodal, and LLM-enhanced systems~\citep{liu2025llm} that jointly optimize accuracy and latency.
In this work, we enhance live streaming recommendation by leveraging generative recommendation techniques.

\section{Preliminaries}

\paratitle{Problem Formulation.}
Let $\mathcal{U}$, $\mathcal{S}$, and $\mathcal{V}$ denote the sets of users, streamers, and currently broadcasting live rooms, respectively. 
Each user $u \in \mathcal{U}$ is described by 
her profiles
$f_u$ (e.g., ID, gender, age) and sequential features $f_h$ (e.g., watched live rooms, interacted products).  
Each streamer $s \in \mathcal{S}$ has features $f_s$ (e.g., follower count, past broadcasts),  
and each live room $v \in \mathcal{V}$ has features $f_v$ (e.g., viewer count, multimodal content).  
We also have user–live interaction features $c_{u,v}$ (e.g., user likes, orders, etc.).
Our objective is to recommend the top-$k$ live rooms 
from $\mathcal{V}$ given these~features.

\paratitle{Pre-Ranking.}
Our method targets the \emph{pre-ranking} stage, which lies between retrieval and ranking in the recommendation pipeline.  
In retrieval, a two-tower model encodes the user and candidate item separately and computes similarity via dot product. 
In ranking, a single-tower model processes raw user and item features via a complex interaction network.
In pre-ranking, the user and item are first encoded by a two-tower architecture, then fed into a lightweight interaction network. 
Formally:
\begin{small}
\begin{equation}
\begin{aligned}
\text{Retrieval:} \quad &\hat{y}_{u,v} = \mathrm{Enc}_u(\mathbf{u}) \cdot \mathrm{Enc}_v(\mathbf{v}), \\
\text{Pre-Ranking:} \quad &\hat{y}_{u,v} = \mathrm{Net}_{\mathrm{pre\text{-}rank}}\big(
        \mathrm{Enc}_u(\mathbf{u}), \mathrm{Enc}_v(\mathbf{v})
    \big), \\
\text{Ranking:} \quad &\hat{y}_{u,v} = \mathrm{Net}_{\mathrm{rank}}\big(\mathbf{u}, \mathbf{v}\big),
\end{aligned}
\end{equation}
\end{small}
where $\mathrm{Enc}_u$ and $\mathrm{Enc}_v$ are user and item encoders,  
$\mathrm{Net}_{\mathrm{pre\text{-}rank}}$ is a lightweight interaction network (e.g., MLP), and $\mathrm{Net}_{\mathrm{rank}}$ is a complex feature-interaction network.  
$\hat{y}_{u,v}$ is the predicted preference score of $u$ for $v$.

In our method, $\mathrm{Enc}_u$ serves primarily as a generative module, computed only once for each user, and is responsible for generating both static and dynamic SIDs. 
$\mathrm{Net}_{\mathrm{pre\text{-}rank}}$ is a lightweight discriminative module to capture user--streamer interaction information, and needs to be computed for each candidate live room.

\section{\ourname Framework}
This section presents our proposed method, \ourname. 
We first describe the construction of static and dynamic Semantic IDs (\S~\ref{sec:live_room_tokenize}), 
followed by an overview of our hybrid generative–discriminative architecture (\S~\ref{sec:architecture}). 
We then detail the training procedure of \ourname (\S~\ref{sec:training}) and its inference process (\S~\ref{sec:inference}).

\subsection{Live Room Tokenization}
\label{sec:live_room_tokenize}
Due to the real-time variability of live rooms, the static semantic ID (SID) commonly used in short video scenarios~\citep{OneRecV1} cannot fully capture their instantaneous dynamics. 
We address this limitation by assigning each live room two complementary identifiers: a \textbf{Static SID} (\S~\ref{sec:static_sid}) derived from historical multimodal segments, and a \textbf{Dynamic SID} (\S~\ref{sec:dynamic_sid}) generated from the current real-time representation. 
Together, they provide a more complete characterization of live room information.

\begin{figure*}[t]
\centering
\includegraphics[width=0.95\textwidth]{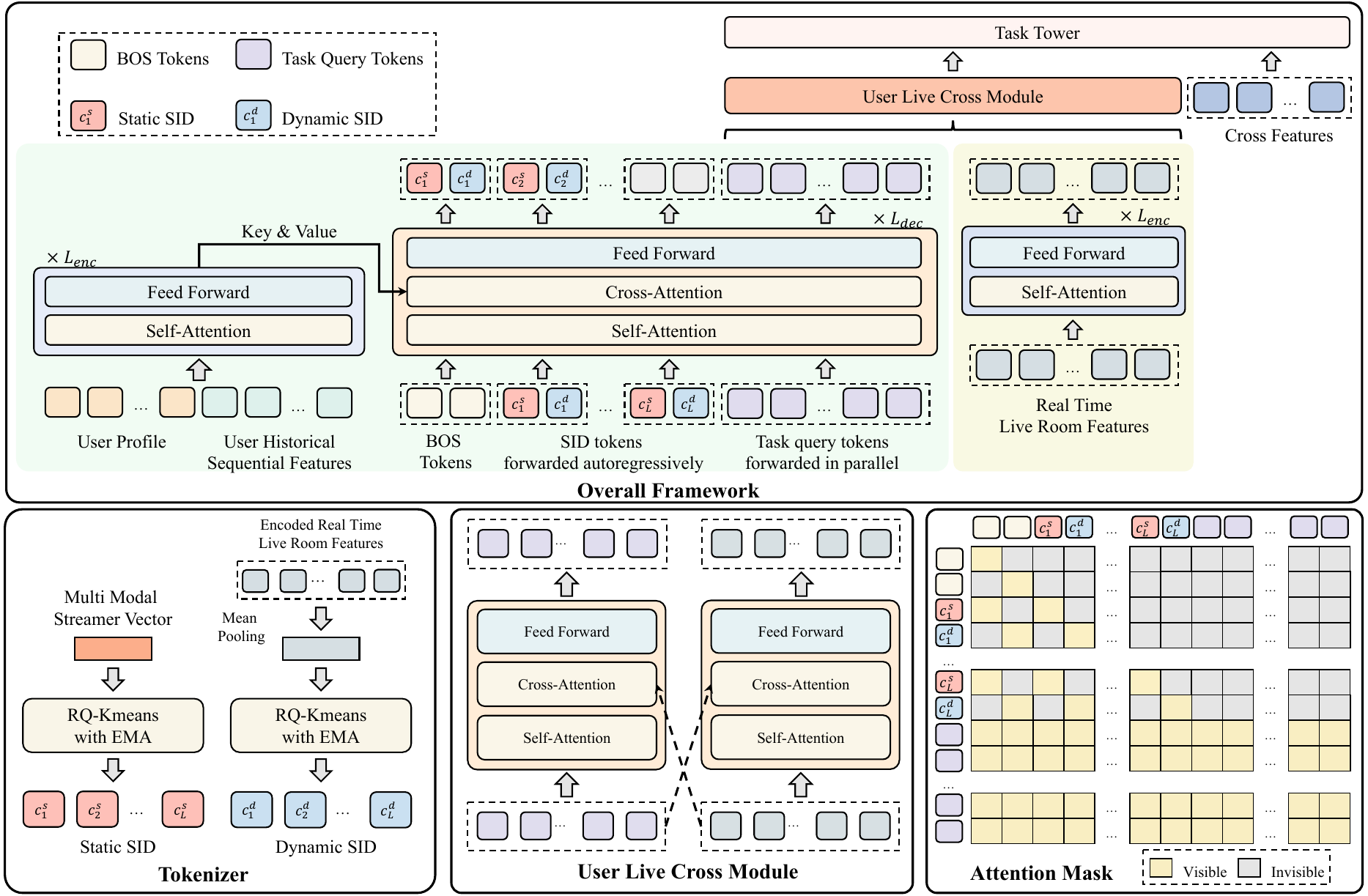}
\vspace{-8px}
\caption{
Overall architecture of \ourname.  
The top panel illustrates the \textit{overall framework}, comprising the user-side SID generation module and the final discriminative module.  
The bottom panels detail: (1) the tokenizer for constructing static and dynamic SIDs;  
(2) the User–Live Cross Module for feature interaction between users and live rooms; and  
(3) the attention mask design in the user-side decoder.
}
\label{fig:method}
\vspace{-0.5cm}
\end{figure*}

\subsubsection{Static Semantic ID}
\label{sec:static_sid}
The static semantic ID (SID) is designed to capture relatively stable characteristics of a streamer that do not vary with the current live content. It is derived from a multimodal vector constructed using the streamer’s historical segments.

\textbf{Multimodal Streamer Vector.} 
We encode multimodal content (e.g., image, audio) from the most recent $M$ days of historical live room segments. 
For each day, $N_{\mathrm{seg}}$ segments are sampled, resulting in $M \times N_{\mathrm{seg}}$ segments per streamer. 
Each segment is embedded with a fine-tuned multimodal model to produce $\mathbf{e}_{\mathrm{seg}} \in \mathbb{R}^d$, 
which are concatenated into $\mathbf{E}_{\mathrm{seg}} \in \mathbb{R}^{(M \times N_{\mathrm{seg}}) \times d}$. 
These embeddings are kept frozen during training. 
A transformer encoder processes $\mathbf{E}_{\mathrm{seg}}$, and the average of its hidden states forms the streamer vector:
\begin{equation}
    \mathbf{h}_s = \mathrm{Mean}\big(\mathrm{TrmEnc}(\mathbf{E}_{\mathrm{seg}})\big), \quad \mathbf{h}_s \in \mathbb{R}^d,
\end{equation}
where $\mathrm{TrmEnc}(\cdot)$ denotes the transformer encoder and $\mathrm{Mean}(\cdot)$ computes the mean pooling over its outputs.

\textbf{Incorporating Collaborative Signals.}  
While $\mathbf{h}_s$ contains rich multimodal semantics, it carries limited collaborative information, which is essential for recommendation~\citep{LETTER}. 
To enrich collaborative signals, we employ the Swing algorithm~\citep{swing} to mine positive streamer pairs and perform streamer-level contrastive learning:
\begin{small}
\begin{equation}
\label{eq:streamer_emb_i2i}
    \mathcal{L}_{\mathrm{SCL}} = 
    \frac{\exp\big(\mathrm{sim}(\mathbf{h}_s, \mathbf{h}_s^{+})\big)}
    {\sum_{\mathbf{h}_s^{-} \in \mathcal{H}_s^{\mathrm{Neg}}} \exp\big(\mathrm{sim}(\mathbf{h}_s, \mathbf{h}_s^{-})\big)},
\end{equation}
\end{small}
where $\mathbf{h}_s^{+}$ is a Swing-mined positive sample, $\mathcal{H}_s^{\mathrm{Neg}}$ is the set of in-batch negatives, and $\mathrm{sim}(\cdot)$ denotes similarity (e.g., dot product or cosine similarity). 
This process yields streamer vectors that jointly encode multimodal and collaborative information.

\textbf{Quantization to Static SID.}  
The trained streamer vector $\mathbf{h}_s$ is quantized via RQ-KMeans~\citep{luo2025qarm} using $L$-level codebooks, producing the static SID:
\begin{equation}
    \mathcal{C}_s = [c_1^s,c_2^s,\ldots,c_L^s],
\end{equation}
where $c_\ell^s$ $(\ell = 1,\ldots,L)$ denotes the index of the selected codeword from the $\ell$-th codebook.

\subsubsection{Dynamic Semantic ID}
\label{sec:dynamic_sid}
Given the inherently real-time nature of live room content, the static SID, which encodes only the streamer’s historical broadcast information, cannot fully capture the instantaneous state of a live room. 
To address this limitation, we introduce the dynamic SID to represent real-time content signals.

\textbf{Real-Time Live Room Encoding.} 
The live room encoding is used both to generate the dynamic SID and to support downstream multi-task discriminative prediction (\S~\ref{sec:multi_task_pred}). 

For an actively broadcasting live room $v \in \mathcal{V}$, 
we extract real-time features (e.g., current viewer count, instantaneous segment-level content) 
as $\mathbf{E}_{f_v} \in \mathbb{R}^{N_{f_v} \times d}$, 
where $N_{f_v}$ is the number of real-time feature fields.

To enable multi-task modeling, $\mathbf{E}_{f_v}$ is flattened, projected via an MLP, 
and reshaped into $\widehat{\mathbf{E}}_{f_v} \in \mathbb{R}^{(T \times Q) \times d}$, 
where $T$ is the number of tasks and $Q$ the number of queries per task:
\begin{equation}
\label{eq:live_room_query}
\widehat{\mathbf{E}}_{f_v} =
\mathrm{Reshape}\big(
    \mathrm{MLP}_v(\mathrm{Flatten}(\mathbf{E}_{f_v}))
\big).
\end{equation}
The slice $\widehat{\mathbf{E}}_{f_v, t} \in \mathbb{R}^{Q \times d}$ is then used to learn task-specific representations for each task $t$ (\S~\ref{sec:multi_task_pred}). 

The transformed embeddings are fed into a Transformer encoder, and mean pooling over its hidden states yields the real-time representation:
\begin{equation}
\label{eq:live_real_time_encoder}
\mathbf{H}_v = \mathrm{TrmEnc}(\widehat{\mathbf{E}}_{f_v}), \quad
\mathbf{h}_v = \mathrm{Mean}(\mathbf{H}_v) \in \mathbb{R}^d.
\end{equation}
We \textbf{warm-start} the live room encoder parameters from a pre-trained live streaming recommendation model 
to leverage prior collaborative and content knowledge, 
facilitating faster convergence and better representation quality.

\textbf{Quantization to Dynamic SID.} 
The real-time live room vector $\mathbf{h}_v$ is quantized via RQ-KMeans~\citep{luo2025qarm} with $L$-level codebooks to produce the dynamic SID:
\begin{equation}
    \mathcal{C}_d = [c_1^d, c_2^d, \ldots, c_L^d],
\end{equation}
where $c_\ell^d \ (\ell = 1,\ldots,L)$ denotes the index of the selected codeword from the $\ell$-th codebook.

\subsubsection{Update of Codebook}
\label{sec:update_codebook}
As discussed in \S~\ref{sec:dynamic_sid}, the real-time live room vectors vary continuously, making a fixed codebook inadequate for capturing rapid changes in representation. To address this, we adopt an \textbf{Exponential Moving Average} (EMA)~\citep{van2017neural,bin2025real} scheme to adaptively update codebook entries.

Let $\{\mathbf{h}_v^{1}, \mathbf{h}_v^{2}, \ldots, \mathbf{h}_v^{B}\}$ denote all real-time vectors in the current batch closest to a given code vector $\mathbf{e}_c \in \mathbb{R}^d$. The EMA update is defined as:
\begin{small}
\begin{equation}
\begin{cases}
N_c^{(t)} = \gamma \, N_c^{(t-1)} + (1-\gamma) \, B, \\
\mathbf{m}_c^{(t)} = \gamma \, \mathbf{m}_c^{(t-1)} + (1-\gamma) \, \sum_{b=1}^B \mathbf{h}_v^{b}, \\
\mathbf{e}_c^{(t)} = \mathbf{m}_c^{(t)} / N_c^{(t)}, 
\end{cases}
\end{equation}
\end{small}
where $N_c$ is the total number of vectors assigned to $\mathbf{e}_c$, $\mathbf{m}_c \in \mathbb{R}^d$ is the sum of these vectors, and $t$ denotes the $t$-th update step. The smoothing factor $\gamma \in (0,1)$ is chosen close to $1$ (e.g., $\gamma = 0.99$) to ensure stable updates.

For multimodal streamer vectors in \S~\ref{sec:static_sid}, updates occur less frequently and the corresponding codebook remains relatively stable. In this case, the EMA update is applied only when the underlying vectors are refreshed.

\subsection{\ourname Architecture}
\label{sec:architecture}
\ourname is a hybrid generative-discriminative architecture, as shown in Figure~\ref{fig:method}. 
For the generative module, we employ an encoder--decoder structure to generate the static and dynamic SIDs obtained in \S~\ref{sec:live_room_tokenize}. 
For the discriminative module, to model user--streamer interaction information (e.g., likes, orders) as cross features, 
we allocate to each task a set of learnable queries that extract task-relevant information from the generated SIDs for the final discriminative prediction. 
This design enables the discriminative module to incorporate rich cross features while simultaneously leveraging the semantic information conveyed by the SIDs.

\subsubsection{User Encoder}
\label{sec:user_encoder}
We represent each user with her profiles (e.g., user ID, age, gender) 
$\mathbf{E}_{f_u} \in \mathbb{R}^{N_{f_u} \times d}$ and historical interaction sequence features 
(e.g., previously watched live rooms) $\mathbf{E}_{f_h} \in \mathbb{R}^{N_{f_h} \times d}$. 
Here, $N_{f_u}$ denotes the number of static feature fields, and $N_{f_h}$ denotes the total length of all historical sequences. 

The static and historical feature embeddings are concatenated and augmented with positional encodings, followed by a Transformer encoder:
\begin{equation}
\begin{aligned}
\mathbf{E}_u = [\mathbf{E}_{f_u}; \mathbf{E}_{f_h}] \in &~\mathbb{R}^{(N_{f_u} + N_{f_h}) \times d},\quad
\widehat{\mathbf{E}}_u = \mathbf{E}_u + \mathbf{P}_u, \\
\mathbf{H}_{\mathrm{Enc}} &= \mathrm{TrmEnc}(\widehat{\mathbf{E}}_u),
\end{aligned}
\end{equation}
where $\mathbf{P}_u \in \mathbb{R}^{(N_{f_u} + N_{f_h}) \times d}$ denotes the learnable positional embeddings, and $\mathrm{TrmEnc}(\cdot)$ is a Transformer encoder with $L_{\mathrm{enc}}$ layers. 
The output $\mathbf{H}_{\mathrm{Enc}}$ is used in the subsequent decoder module.

\subsubsection{Semantic ID Decoder}
\label{sec:sid_decoder}
For the decoder input, we prepend two begin-of-sequence (\texttt{[BOS]}) tokens for the static and dynamic SIDs, followed by interleaved SID tokens and a set of learnable queries for downstream multi-task objectives:
\begin{small}
\begin{equation}
\label{eq:dec_inp}
\begin{aligned}
    \mathrm{DecInp} = \big\{ & [\texttt{[BOS]}_s, \texttt{[BOS]}_d], 
             [c_1^s, c_1^d], \ldots, [c_L^s, c_L^d], \\
           & [q^{1}_1, \ldots, q^{1}_{Q}, \ldots, q^{T}_1, \ldots, q^{T}_{Q}] \big\},
\end{aligned}
\end{equation}
\end{small}
where \texttt{[BOS]}$_s$ and \texttt{[BOS]}$_d$ denote the static and dynamic SID \texttt{[BOS]} tokens, $Q$ is the number of queries per task, and $T$ is the number of tasks in the final multi-task stage.  
Static and dynamic SIDs are organized \emph{interleaved}, enabling their simultaneous generation at each level (e.g., $[c_i^s, c_i^d]$ at level $i$) and avoiding the doubled generation time of naive concatenation.

Dedicated embedding tables are maintained for all SID tokens, \texttt{[BOS]} tokens, and learnable queries. The decoder input embeddings are obtained via table lookup:
\begin{small}
\begin{equation}
\begin{aligned}
\mathbf{E}_{\mathrm{DecInp}} = [&\mathbf{e}_{\texttt{[BOS]}_s}; \mathbf{e}_{\texttt{[BOS]}_d}; \mathbf{e}_{c_1^s}; \mathbf{e}_{c_1^d}; \ldots; \mathbf{e}_{c_L^s}; \mathbf{e}_{c_L^d}; \\
&\mathbf{e}_{q^{1}_1}; \ldots; \mathbf{e}_{q^{1}_{Q}}; \ldots; \mathbf{e}_{q^{T}_1}; \ldots; \mathbf{e}_{q^{T}_{Q}} ],
\end{aligned}
\end{equation}
\end{small}
where $\mathbf{E}_{\mathrm{DecInp}} \in \mathbb{R}^{(2 + 2L + TQ) \times d}$ is the decoder input embeddings.

Positional encodings are added and the sequence is passed to a Transformer decoder:
\begin{equation}
\begin{aligned}
\widehat{\mathbf{E}}_{\mathrm{DecInp}} &= \mathbf{E}_{\mathrm{DecInp}} + \mathbf{P}_{\mathrm{Dec}}, \\
\mathbf{H}_{\mathrm{Dec}} &= \mathrm{TrmDec}\big(\widehat{\mathbf{E}}_{\mathrm{DecInp}}, \mathbf{H}_{\mathrm{Enc}}, \mathbf{H}_{\mathrm{Enc}}\big),
\end{aligned}
\end{equation}
where $\mathbf{P}_{\mathrm{Dec}} \in \mathbb{R}^{(2 + 2L + TQ) \times d}$ is the learnable positional embedding, and $\mathrm{TrmDec}(\cdot)$ denotes a Transformer decoder with $L_{\mathrm{dec}}$ layers that uses encoder output $\mathbf{H}_{\mathrm{Enc}}$ (\S~\ref{sec:user_encoder}) as both key and value.

\textbf{Attention Mask.}  
We design the decoder self-attention mask as shown in Figure~\ref{fig:method}.  
All task queries are propagated jointly within a single forward pass and are mutually visible, while also attending to all preceding SID tokens.  
For SID positions, static and dynamic SIDs attend only to their respective preceding tokens (e.g., a static SID only attends to previously generated static SIDs), preserving information separation between the two types.

\subsubsection{User–Live Cross Module}
\label{sec:user_live_cross_attn}
We take the last $T \times Q$ hidden states from $\mathbf{H}_{\mathrm{Dec}}$ as 
$\mathbf{H}_{u} \in \mathbb{R}^{(T \times Q) \times d}$, corresponding to all learnable queries. 
These states encode task-relevant information extracted from both SID representations and user features.

To model user–live interactions, we perform cross-attention between $\mathbf{H}_{u}$ and the live room representation 
$\mathbf{H}_{v}$ from Eq.~\eqref{eq:live_real_time_encoder}. 
For the live room branch:
\begin{small}
\begin{equation}
\label{eq:user_live_cross}
\begin{aligned}
    \widehat{\mathbf{H}}_{v} &= \mathrm{SelfAttn}(\mathbf{H}_{v}), \\
    \widehat{\mathbf{H}}_{v} &= \mathrm{FFN}\Big(
        \mathrm{CrossAttn}\big(\widehat{\mathbf{H}}_{v}, \mathbf{H}_{u}, \mathbf{H}_{u}\big) 
        + \widehat{\mathbf{H}}_{v} \Big),
\end{aligned}
\end{equation}
\end{small}
where $\widehat{\mathbf{H}}_{v} \in \mathbb{R}^{(T \times Q) \times d}$ 
captures live room features enriched with cross-interaction information.  
The same procedure is applied symmetrically to obtain $\widehat{\mathbf{H}}_{u}$ for the user branch.

\subsubsection{Multi-Task Prediction}
\label{sec:multi_task_pred}
Finally, we use the enriched user features $\widehat{\mathbf{H}}_{u}$, 
live room features $\widehat{\mathbf{H}}_{v}$ from \S~\ref{sec:user_live_cross_attn}, 
and the cross features $\mathbf{c}_{u,v} \in \mathbb{R}^d$ for multi-task prediction:
\begin{equation}
\label{eq:multi_task_pred}
\begin{aligned}
    \mathbf{h}_t &= \mathrm{Concat}\big(
        \mathrm{Flatten}(\widehat{\mathbf{H}}_{u,t}),
        \mathrm{Flatten}(\widehat{\mathbf{H}}_{v,t}),
        \mathbf{c}_{u,v}
    \big), \\
    \hat{y}_t &= \mathrm{MLP}_t(\mathbf{h}_t),
\end{aligned}
\end{equation}
where $\mathrm{Flatten}(\cdot)$ converts a matrix to a 1D vector, and 
$\widehat{\mathbf{H}}_{u,t}, \widehat{\mathbf{H}}_{v,t} \in \mathbb{R}^{Q \times d}$ 
are the slices corresponding to task $t$.  
This design enables learning task-specific representations for each prediction objective, 
with $\hat{y}_t$ denoting the prediction score for task $t$.

\subsection{\ourname Training}
\label{sec:training}

\subsubsection{Semantic ID Prediction}
\label{sec:sid_pred}
We use the first $2L$ hidden states from $\mathbf{H}_{\mathrm{Dec}}$ to predict both static and dynamic SIDs. 
The cross-entropy loss is defined as:
\begin{small}
\begin{equation}
\label{eq:sid_ntp}
\mathcal{L}_{\mathrm{NTP}} = -\frac{1}{L} \sum_{i=3}^{2 + 2L}
\begin{cases}
\log \mathrm{P}\left(c_{\lfloor i/2 \rfloor - 1}^d \mid \mathrm{DecInp}_{< i}\right), & i \bmod 2 = 0, \\
\log \mathrm{P}\left(c_{\lfloor i/2 \rfloor}^s \mid \mathrm{DecInp}_{< i}\right), & i \bmod 2 = 1,
\end{cases}
\end{equation}
\end{small}
where $i \bmod 2 = 0$ refers to dynamic SID tokens, and $i \bmod 2 = 1$ refers to static SID tokens.  
The index $i$ starts from $3$ because, according to Eq.~\eqref{eq:dec_inp}, SID tokens occupy positions beginning at the third element.  
The terms $\lfloor i/2 \rfloor - 1$ and $\lfloor i/2 \rfloor$ convert the decoder input index $i$ into the corresponding dynamic and static SID indices, respectively.  
$\mathrm{DecInp}_{< i}$ denotes the partial decoder input preceding position $i$ used as the conditioning context, with 
$\mathrm{DecInp}_{< 3} = [\texttt{[BOS]}_s, \ \texttt{[BOS]}_d]$ at initialization.

\subsubsection{Multi-Task Learning}
The multi-task loss is computed as:
\begin{equation}
\mathcal{L}_{\mathrm{MTL}} = \sum_{t=1}^{T} w_t \, \mathcal{L}_t (y_t, \hat{y}_t),
\end{equation}
where $y_t$ is the ground-truth label for task $t$, 
$\mathcal{L}_t$ is the task-specific loss function 
(e.g., binary cross-entropy for binary classification),  
and $w_t$ is the weight assigned to task $t$.

The overall training objective is:
\begin{equation}
\mathcal{L}_\mathrm{Total} = \mathcal{L}_\mathrm{MTL} 
+ \lambda_{\mathrm{NTP}}\, \mathcal{L}_{\mathrm{NTP}} 
+ \lambda_{\mathrm{Reg}} \|\Theta\|^2_2,
\end{equation}
where $\lambda_{\mathrm{NTP}}$ controls the contribution of the Semantic ID prediction loss,  
$\lambda_{\mathrm{Reg}} \|\Theta\|^2_2$ is the $L_2$ regularization term, 
and $\Theta$ denotes all model parameters.

\subsection{\ourname Inference}
\label{sec:inference}

\subsubsection{Beam Fusion}
\label{sec:beam_fusion}
During training, the next-token prediction (NTP) loss in Eq.~\eqref{eq:sid_ntp} 
is minimized using the teacher forcing strategy~\citep{raffel2020exploring}, 
where ground-truth tokens are provided to the decoder at each step to guide SID generation.

At inference, we employ a beam search strategy to jointly generate static and dynamic SIDs. 
Unlike conventional beam search, our method maintains two independent beam sets, 
each producing $B$ candidate sequences for static and dynamic SIDs, respectively. 
For each static candidate sequence, the score for the $b$-th beam is computed as:
\begin{small}
\begin{equation}
\begin{aligned}
p_b^s &= \sum_{\ell=1}^{L} 
\log \mathrm{P}\!\left( c_{b,\ell}^s \,\middle|\, \texttt{[BOS]}_s,\, c_{b,1}^s,\, \ldots,\, c_{b,\ell-1}^s \right), \\
\hat{p}_b^s &= \exp(p_b^s) \big/ \sum_{b'=1}^{B} \exp(p_{b'}^s),
\end{aligned}
\end{equation}
\end{small}
where $\hat{p}_b^s$ denotes the normalized probability of the $b$-th static SID candidate. 
Normalized probabilities $\hat{p}_b^d$ are obtained analogously for dynamic SIDs.

Each of the $B$ candidate sequences yields a set of hidden states 
$\mathbf{H}_{u,b}$ (\S~\ref{sec:user_live_cross_attn}). 
We compute the fused representation by weighting each $\mathbf{H}_{u,b}$ 
with the average normalized probability of its corresponding static and dynamic candidates:
\begin{small}
\begin{equation}
    \mathbf{H}_u^{\mathrm{Fuse}} = \sum_{b=1}^{B} (\hat{p}_b^s + \hat{p}_b^d) / 2 \cdot \mathbf{H}_{u,b}.
\end{equation}
\end{small}
The aggregated representation $\mathbf{H}_u^{\mathrm{Fuse}}$ is then passed through 
Eq.~\eqref{eq:user_live_cross} and Eq.~\eqref{eq:multi_task_pred} to produce the final predictions.

\subsubsection{Fusion Score}
In the online serving stage, multi-task prediction scores from Eq.~\eqref{eq:multi_task_pred} 
are combined into a single fusion score for pre-ranking:
\begin{equation}
\label{eq:fusion_score}
    y_{\mathrm{Fuse}} = \prod_{t=1}^{T} \hat{y}_t^{\alpha_t},
\end{equation}
where $\alpha_t$ is the weight for task $t$, tuned according to online performance.

\subsection{Discussion}
\label{sec:discussion}

\subsubsection{Time Complexity Analysis}
We analyze the computational complexity of \ourname to evaluate its efficiency.

\noindent\textbf{User Side:}  
The model adopts an encoder–decoder architecture.  
For the encoder, the complexity is $O(N_u^2 d + N_u d^2)$,  
where $N_u = N_{f_u} + N_{f_h}$, with $N_{f_u}$ denoting the number of static features and $N_{f_h}$ the length of the user's historical sequence.  
For the decoder, the complexity is $O(N_d^2 d + N_d d^2 + N_d N_u d + N_u d^2)$,  
where $N_d = 2 + 2L + T Q$, $L$ is the SID length, $T$ the number of tasks, and $Q$ the queries per task.  
Since $T, Q \ll N_u$, the decoder cost is relatively~low.

\noindent\textbf{Live Room Side:}  
Live room features are first mapped to  
$\widehat{\mathbf{E}}_{f_v} \in \mathbb{R}^{(T Q) \times d}$  
(Eq.~\eqref{eq:live_room_query}) and then processed by a Transformer encoder,  
with complexity $O(N_q^2 d + N_q d^2)$, where $N_q = T Q$.

\noindent\textbf{User–Live Cross Module:}  
Operating on $\mathbf{H}_u, \mathbf{H}_v \in \mathbb{R}^{(T Q) \times d}$  
(\S~\ref{sec:user_live_cross_attn}),  
its cost is $O(N_q^2 d + N_q d^2)$,  
which is minor compared to the user-side encoder given $T, Q \ll N_u$.
\section{Offline Experiments}
To validate the effectiveness of \ourname, we first conducted extensive offline experiments.

\subsection{Experimental Setup}

\subsubsection{Dataset}
To the best of our knowledge, there is no publicly available dataset that simultaneously contains real-time features of the live room and multimodal information from historical live room segments of the streamer~\citep{qu2025kuailive}. Therefore, we collect commercial live streaming data from Taobao\footnote{\url{https://taolive.taobao.com/}} to conduct offline experiments.

We collected one week of interaction data between users and streamers, 
amounting to approximately 1 billion (1B) interaction records.  
Each record contains user watch time, whether the user placed an order, and whether the user interacted with the streamer, among other behaviors.

\subsubsection{Baseline Methods}
We compared \ourname with the following representative baselines:

\textbf{DLRM}~(Taobao Live Streaming Baseline):  
A production pre-ranking model with a \textit{dual-tower} architecture and a \textit{feature-crossing} module.  
The user tower applies self-attention to historical behaviors and target-attention with queries from static features;  
the live room tower encodes its features similarly.  
Top-layer feature crossing concatenates user and live room representations with cross features, which are fed into MLPs with personalized gating via ppnet~\citep{chang2023pepnet} for multi-task prediction.

\textbf{SASRec}~\citep{SASREC} adopts a unidirectional Transformer decoder to model user interaction histories for next-item prediction. 
\textbf{ReaRec}~\citep{ReaRec} performs multi-step latent reasoning over item sequences to derive user representations. 
\textbf{HSTU}~\citep{HSTU} is a high-performance self-attention encoder for industrial-scale recommendation, and serves as a backbone for generative recommenders.

\textbf{\emph{Adaptation to Pre-ranking.}}  
Our method operates in pre-ranking, while several baselines (e.g., SASRec, ReaRec, HSTU) target retrieval, making direct comparison inappropriate.  
For fairness, we adapt them by using their backbones to encode user and live room features; the resulting representations and cross features are concatenated and fed into MLPs for multi-task prediction.
For more details about the baselines, please refer to Appendix~\ref{appendix:baseline}.

\subsubsection{Evaluation Metrics}
Our model is evaluated in the pre-ranking stage. 
Following prior work~\citep{OnePiece,Rankmixer}, we report AUC and GAUC for each task. 
The evaluation tasks include user watch time and user order (whether the user places an order). 
Watch time is discretized into two buckets: \textit{watch30} and \textit{watch200}, which indicate whether a user’s watch duration exceeds 30 or 200 seconds, respectively.

\begin{table}[!t]
\centering
\caption{
Offline performance of different methods in the pre-ranking stage for our scenario. 
Best and second-best results are highlighted in bold and underlined, respectively, with statistically significant improvements over the latter.
}
\vspace{-8px}
\label{tab:offline_result}
\renewcommand{\arraystretch}{1.2}
\resizebox{1.0\columnwidth}{!}{
\begin{tabular}
{l
cccccc
}
\toprule
\multirow{2}{*}{Model} 
&\multicolumn{2}{c}{Watch30}
&\multicolumn{2}{c}{Watch200} 
&\multicolumn{2}{c}{Order} 
\\
\cmidrule(l){2-3}  \cmidrule(l){4-5} 
\cmidrule(l){6-7}  
&AUC &GAUC 
&AUC &GAUC 
&AUC &GAUC 
\\
\midrule
DLRM & {0.7610} & \underline{0.6892} & \underline{0.8229} & \textbf{0.7288} & \underline{0.8312} & \underline{0.6727} \\
SASRec & 0.7581 & 0.6781 & 0.8154 & 0.7105 & 0.8205 & 0.6689 \\
ReaRec & 0.7594 & 0.6806 & 0.8167 & 0.7145 & 0.8216 & 0.6689 \\
HSTU & \underline{0.7613} & 0.6823 & 0.8192 & 0.7156 & 0.8249 & 0.6710 \\
\midrule
\textbf{\ourname} & \textbf{0.7692} & \textbf{0.6956} & \textbf{0.8255} & \underline{0.7281} & \textbf{0.8358} & \textbf{0.6946} \\
\bottomrule
\end{tabular}
} 
\vspace{-0.3cm}
\end{table}

\subsection{Overall Performance}
Table~\ref{tab:offline_result} reports the offline performance of \ourname and baseline models on industrial datasets. 
From the results, we observe that:

\noindent\textbf{$\bullet$} First, \ourname achieves the best overall performance. 
Compared with our strong online baseline DLRM and other representative recommendation models, it consistently delivers superior results. 
This validates the effectiveness of combining dynamic SIDs with a hybrid generative–discriminative architecture.

\noindent\textbf{$\bullet$} Second, DLRM proves to be a very strong baseline. 
In most cases, it outperforms other representative recommendation models (e.g., HSTU), 
confirming its high performance level and the challenge of further improvement upon such a strong baseline.

\noindent\textbf{$\bullet$}
Finally, we observe that stronger backbones consistently deliver better performance.  
SASRec, based solely on a transformer decoder, performs relatively poorly, whereas methods incorporating latent reasoning (ReaRec) or leveraging a more powerful backbone (HSTU) achieve notably superior results.  
These findings underscore the critical role of the backbone architecture.

\begin{table}[!t]
\centering
\caption{
Ablation results for \ourname, where ``w/o'' indicates that the corresponding module has been removed 
from the~model.
}
\vspace{-8px}
\label{tab:ablation_study}
\renewcommand{\arraystretch}{1.2}
\resizebox{1.0\columnwidth}{!}{
\begin{tabular}
{c
cccccc
}
\toprule
\multirow{2}{*}{Model} 
&\multicolumn{2}{c}{Watch30}
&\multicolumn{2}{c}{Watch200} 
&\multicolumn{2}{c}{Order} 
\\
\cmidrule(l){2-3}  \cmidrule(l){4-5} 
\cmidrule(l){6-7}  
&AUC &GAUC 
&AUC &GAUC 
&AUC &GAUC 
\\
\midrule
\textbf{\ourname} & \textbf{0.7692} & \textbf{0.6956} & \textbf{0.8255} & \textbf{0.7281} & \textbf{0.8358} & \textbf{0.6946} \\
\midrule
w/o SID & 0.7602 & 0.6808 & 0.8179 & 0.7156 & 0.8245 & 0.6773 \\
\hdashline
w/o Dynamic SID & 0.7637 & 0.6861 & 0.8214 & 0.7205 & 0.8280 & 0.6798 \\
\hdashline
w/o Task Queries & 0.7676 & 0.6929 & 0.8246 & 0.7266 & 0.8344 & 0.6960 \\
\hdashline
w/o Cross Features & 0.7593 & 0.6800 & 0.8149 & 0.7100 & 0.8269 & 0.6826 \\
\hdashline
\makecell[c]{w/o User-Live \\ Cross Module} & 0.7675 & 0.6910 & 0.8248 & 0.7253 & 0.8341 & 0.6878 \\
\bottomrule
\end{tabular}
} 
\vspace{-0.3cm}
\end{table}

\begin{table}[!t]
\centering
\caption{
Dynamic SIDs of the same streamer at different timestamps with corresponding live room engagement statistics, 
including views and total watch time within 5 and 30 minutes (in minutes). 
For aggregation, records sharing the same first two levels of the dynamic SID are summed. 
The static SID of this streamer remains (1512, 646, 631) for all timestamps.
}
\vspace{-8px}
\label{tab:dynamic_sid_case}
\renewcommand{\arraystretch}{1.2}
\resizebox{1.0\columnwidth}{!}{
\begin{tabular}
{ccccc}
\toprule
Dynamic SID & \makecell[c]{Views \\ (5 min)} & \makecell[c]{Views \\ (30 min)} & \makecell[c]{Total Watch Time \\ (5 min, min)} & \makecell[c]{Total Watch Time \\ (30 min, min)} \\ 
\midrule
(111, 1559, *) & 92 & 193 & 24   & 84   \\
(111, 1551, *) & 509 & 3,769 & 420  & 3,755 \\
(111, 1194, *) & 773 & 4,441 & 725  & 4,172 \\
\bottomrule
\end{tabular}
} 
\vspace{-0.5cm}
\end{table}

\subsection{Ablation Study}
In this section, we perform ablation studies to assess the contribution of each module in \ourname, as summarized in Table~\ref{tab:ablation_study}.

\noindent\textbf{$\bullet$ Effects of Semantic IDs.}  
Removing SIDs leads to clear performance degradation, validating their effectiveness. 
Similarly, removing dynamic SIDs results in a drop in performance, confirming their utility.
Moreover, the ``w/o Dynamic SID'' setting performs better than ``w/o SID'', which indicates that static SIDs are also effective.
Overall, both static and dynamic SIDs contribute positively to the model's performance.

\noindent\textbf{$\bullet$ Effects of Task Queries.}  
Removing task queries, i.e., directly using the decoder’s last hidden state for prediction, leads to degraded performance.  
This confirms their effectiveness in extracting task-relevant information from SIDs and user features.

\noindent\textbf{$\bullet$ Effects of Features Crossing.}  
Removing either the cross features or the user–live cross module (\S~\ref{sec:user_live_cross_attn}) leads to performance degradation, validating the importance of cross features between users and streamers (i.e., interaction information).  
This also indicates that purely generative SID methods are insufficient for our~scenario.

\subsection{Experimental Analysis}

\subsubsection{Case Study of Dynamic SID}
We conduct a case study to examine the effectiveness of dynamic SIDs, as shown in Table~\ref{tab:dynamic_sid_case}. 
We analyze the dynamic SIDs and corresponding live room engagement statistics for the same streamer at different timestamps within a single day. 
The results show that as the live room's popularity changes, the dynamic SIDs also vary accordingly. 
Moreover, different SID prefixes reflect variations in factors such as the real-time popularity level of the same live room. 
This observation confirms that dynamic SIDs can effectively capture temporal changes in live streaming scenarios.
For a case study on static SIDs, please refer to Appendix~\ref{appendix:static_sid_case}.

\subsubsection{Task Query Visualization}
\begin{figure}[t]
\centering
\includegraphics[width=0.9\columnwidth]{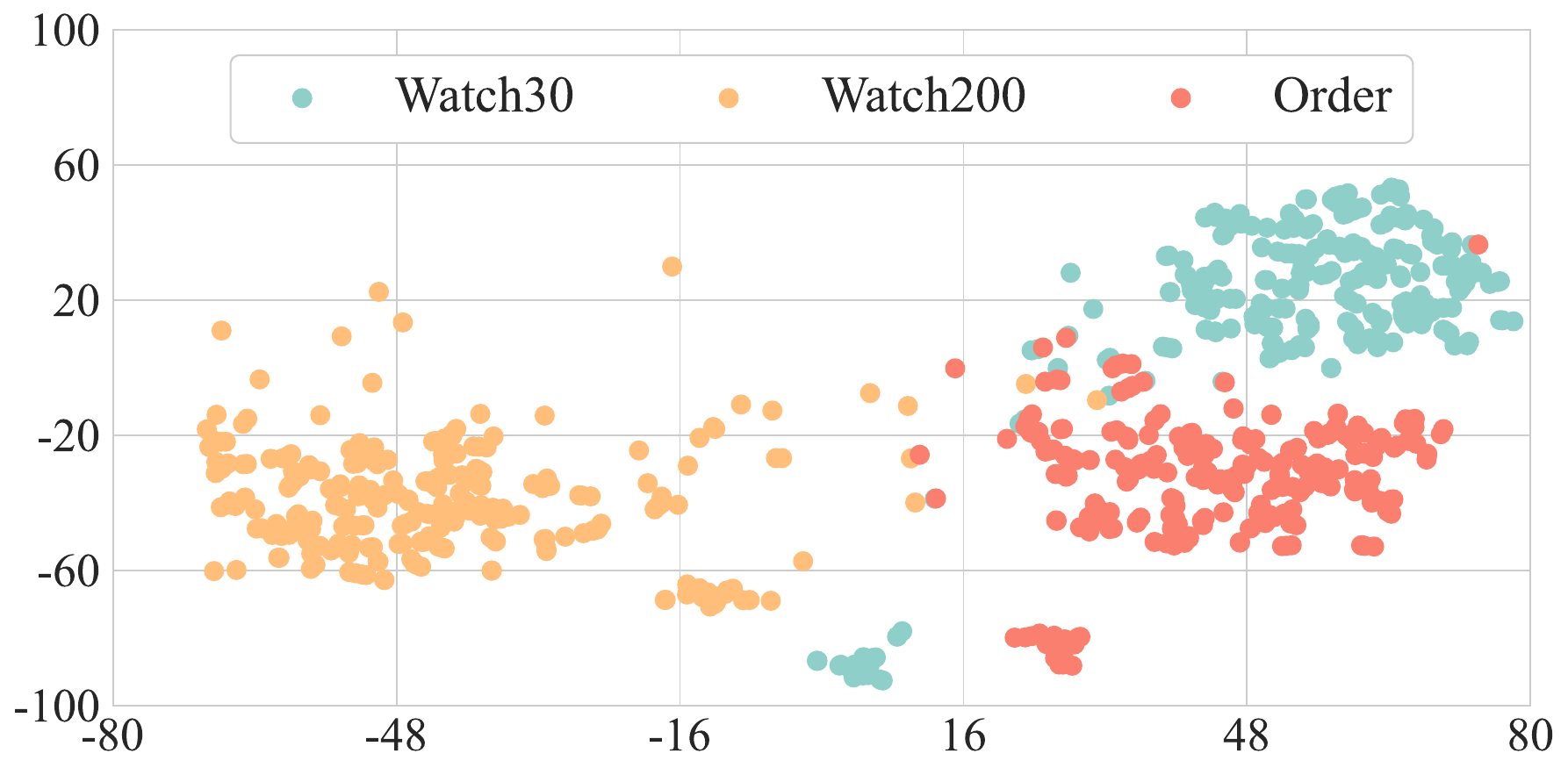}
\vspace{-8px}
\caption{
The t-SNE visualization of the task-specific query extraction parts in $\mathbf{H}_{u}$ (\S~\ref{sec:user_live_cross_attn}). 
}
\label{fig:query_tsne}
\vspace{-0.3cm}
\end{figure}

We apply t-SNE~\citep{van2008visualizing} to reduce the dimensionality of the $\mathbf{H}_{u}$ extracted by different task queries in \S~\ref{sec:user_live_cross_attn}, and then visualize the resulting representations, as shown in Figure~\ref{fig:query_tsne}. 
Different task queries yield user representations that are distributed in distinct clusters. 
This indicates, on one hand, that different tasks require different types of information, and on the other hand, that task queries successfully learn diverse information without collapsing, confirming their effectiveness.
Further analysis of the task query can be found in Appendix~\ref{appendix:task_query}.

\begin{figure}[t]
\centering
\includegraphics[width=0.95\columnwidth]{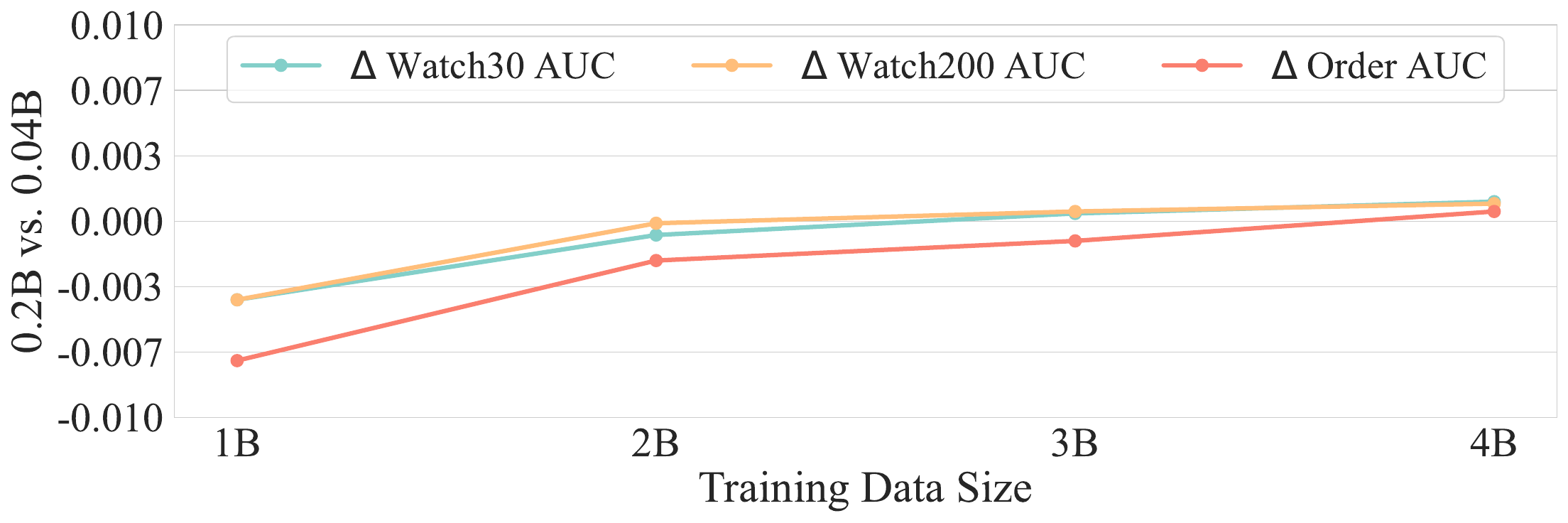}
\vspace{-8px}
\caption{
AUC differences between \ourname~(0.2B) and \ourname~(0.04B) across various targets as the training data size~increases.
}
\label{fig:scale_diff}
\vspace{-0.5cm}
\end{figure}

\subsubsection{Scaling Analysis}
We investigate the impact of increasing model parameters on the AUC of each target, as illustrated in Figure~\ref{fig:scale_diff}.  
In particular, we examine how the AUC difference between the 0.2B-parameter model and the 0.04B-parameter model varies with the size of the training data.  
We observe that with limited training data, the 0.2B model underperforms the 0.04B model.  
However, as the amount of training data increases, the 0.2B model gradually surpasses the 0.04B model, indicating that larger models converge more slowly and require more training data, yet possess a higher performance ceiling.
\section{Online Experiments}
We further validate the effectiveness of \ourname through real-world online experiments.

\subsection{Experimental Setup}

\subsubsection{Evaluation Protocol}
We conduct an
online A/B testing experiment to compare the performance of \ourname against the deployed pre-ranking baseline model.

\subsubsection{Evaluation Metrics}
On our live streaming platform, we assess model performance in the online setting using both business outcomes and user-engagement indicators.

\textit{Core metrics} include:
\textbf{Watch Time}, which is the total time users spend watching the live broadcast;
\textbf{Watch Count}, which is the number of viewing sessions initiated by users;
\textbf{GMV} (\textit{Gross Merchandise Volume}), which is the total value of transactions completed during the session; and
\textbf{Order Count}, which is the number of completed purchases.
\textit{Interaction metrics} comprise:
\textbf{Follower Count}, which is the number of users who follow the streamer; and
\textbf{Interaction Count}, which is the number of unique users engaging in activities such as likes and comments.

We also report the \textbf{Hit Rate} for different objectives, defined as the proportion of cases where a positive instance of the target appears within the model’s final top-$k$ ranked list.

\begin{table}[!t]
\centering
\caption{
A/B testing results in our online live streaming platform.
All metric improvements are statistically significant in our~scenario.
}
\vspace{-8px}
\label{tab:online_result}
\renewcommand{\arraystretch}{1.2}
\resizebox{1.0\columnwidth}{!}{
\begin{tabular}
{ccccccc}
\toprule
\multirow{2}{*}{Model}
&\multicolumn{4}{c}{Core Metrics} 
&\multicolumn{2}{c}{Interaction Metrics} \\
\cmidrule(l){2-5} \cmidrule(l){6-7} 
&\makecell[c]{Watch \\ Time} &\makecell[c]{Watch \\ Count} &GMV &\makecell[c]{Order \\ Count}
&\makecell[c]{Follower \\ Count}
&\makecell[c]{Interaction \\ Count}
\\ 
\midrule
\makecell[c]{\textbf{\ourname} \\ vs. DRLM} &+3.38\% &+1.89\% &+0.72\% &+2.36\%  &+3.12\%   &+2.92\%\\
\bottomrule
\end{tabular}
} 
\vspace{-0.3cm}
\end{table}

\begin{table}[!t]
\centering
\caption{
Online Hit Rate. For each objective, the Hit Rate is computed by checking whether a positive instance of the corresponding target appears in the model's final top-$k$ ranked~list.
}
\vspace{-8px}
\label{tab:online_hit_rate}
\renewcommand{\arraystretch}{1.2}
\resizebox{0.9\columnwidth}{!}{
\begin{tabular}
{lcccc}
\toprule
Model
&Watch30 
&Watch200 &Order &Interact \\ 
\midrule
DLRM &0.7172 
&0.6200 &0.6065 &0.6350 \\
\textbf{\ourname} &\textbf{0.7438} 
&\textbf{0.6428} &\textbf{0.6274}  &\textbf{0.6580} \\
\hline
Improv. &+0.0266 
&+0.0228 &+0.0208 &+0.0230 \\
\bottomrule
\end{tabular}
} 
\vspace{-0.3cm}
\end{table}

\begin{table}[!t]
\centering
\caption{
Online serving efficiency. Latency is expressed as a percentage relative to the baseline DLRM ($100\%$). 
Partial Run refers to the optimization in \S~\ref{sec:online_serve}.
}
\vspace{-8px}
\label{tab:online_efficiency}
\renewcommand{\arraystretch}{1.2}
\resizebox{0.95\columnwidth}{!}{
\begin{tabular}
{lcccc}
\toprule
Model
&\#Param Dense &FLOPs & Latency &\makecell[c]{Latency \\ (Partial Run)} \\ 
\midrule
DLRM &3M  &0.9T &100\%  &100\% \\
\textbf{\ourname} &0.04B &15T  &104.41\% &101.33\%  \\
\bottomrule
\end{tabular}
} 
\vspace{-0.3cm}
\end{table}

\subsubsection{Online Serving}
\label{sec:online_serve}
We incorporate engineering optimizations to improve online serving efficiency:

\noindent\textbf{$\bullet$ Partial Run.}  
Since data preparation constitutes a substantial portion of latency in online recommender systems,  
we start generative module computation in advance before candidate live room data preparation completes.  
This optimization effectively shortens service response time.
For more details about online serving, please refer to Appendix~\ref{appendix:online_serve}.

\subsection{Online Results}

\subsubsection{Overall Performance}
Tables~\ref{tab:online_result} and~\ref{tab:online_hit_rate} report the online A/B testing performance of \ourname against the DLRM baseline.  
The results show consistent and statistically significant improvements across all metrics:

\noindent\textbf{$\bullet $}~\textbf{Core metrics:} 
\ourname increases Watch Time by $+3.38\%$ and Watch Count by $+1.89\%$, indicating longer viewing durations and more viewing sessions. GMV and Order Count are improved by $+0.72\%$ and $+2.36\%$, demonstrating significant gains in monetization.

\noindent\textbf{$\bullet $}~\textbf{Interaction metrics:} 
Follower count increases by $+3.12\%$ and interaction count by $+2.92\%$, 
indicating that \ourname's recommendations better connect viewers with preferred streamers 
and encourage more interactive behaviors, such as likes and comments.

\noindent\textbf{$\bullet$ Online Hit Rate:}  
Across all objectives, \ourname achieves higher Hit Rates, with gains of $+0.0266$, and $+0.0228$ for Watch30 and Watch200, respectively, and $+0.0208$ and $+0.0230$ for Order and Interact.  
These improvements demonstrate better alignment between predicted rankings and actual user behaviors.

Overall, the A/B testing results confirm that \ourname robustly outperforms a strong industrial baseline, boosting engagement, conversion, and interaction in real-world live streaming scenarios.

\subsubsection{Efficiency Analysis}
We analyze the computational overhead of \ourname compared with DLRM, as shown in Table~\ref{tab:online_efficiency}. 
\ourname contains 0.04B dense parameters and requires 15T FLOPs, which is substantially larger than DLRM's 3M parameters and 0.9T FLOPs, indicating a higher utilization of computational resources. 
Despite the increased model complexity, \ourname incurs only a small average latency overhead, approximately $+4.41\%$ without optimizations, and merely $+1.33\%$ with Partial Run enabled.
Further analysis can be found in Appendix~\ref{appendix:efficieny}.

\section{Conclusion}
\label{sec:Conclusion}
We proposed \ourname, a hybrid generative–discriminative recommendation model for live streaming scenarios. 
It combines static and dynamic Semantic IDs to represent multimodal streamer content and real-time dynamics, and uses task queries with cross-feature fusion to explicitly model user–streamer interactions. 
Experiments on industrial datasets show consistent improvements over strong baselines, while online A/B testing demonstrates notable gains in watch time, gross merchandise value (GMV), and interaction metrics. 
With engineering optimization, these improvements come with negligible end-to-end latency overhead, confirming the effectiveness of \ourname in large-scale real-world deployments.

\bibliographystyle{ACM-Reference-Format}
\bibliography{bibtex}

\appendix

\section{Experimental Details}

\subsection{Baselines}
\label{appendix:baseline}
We provide more details about the baselines:

\textbf{DLRM}~(Production Baseline in the Taobao Live Streaming Scenario):  
The pre-ranking baseline deployed in Taobao Live Streaming is a well-optimized production model.  
Its core architecture adopts a \textit{dual-tower structure} followed by a \textit{feature-crossing module}.  
In the \textit{dual-tower component}, the user side and live room side are encoded independently.  
On the user side, the encoding process incorporates a self-attention mechanism over the user's historical behavior sequence. The resulting representation is further refined via a target-attention module, in which the query is derived from the user's static features and applied to the historical sequence.  
Similarly, the live room side is encoded using its corresponding features. 
At the \textit{top layer}, a lightweight feature-crossing module concatenates the user representation, live room representation, and cross features. The combined features are then passed to multi-layer perceptrons (MLPs) to produce outputs for multiple prediction tasks.  
The MLPs employ personalized gating control through the ppnet~\citep{chang2023pepnet} module, enabling task-specific adaptation.

\textbf{SASRec}~\citep{SASREC} employs a unidirectional Transformer decoder to encode a user's historical interactions and predict the next item. 

\textbf{ReaRec}~\citep{ReaRec} is a representative reasoning-augmented recommendation framework that models user representations through multi-step latent reasoning over item interaction sequences.

\textbf{HSTU}~\citep{HSTU}
is a high-performance self-attention encoder 
architecture tailored for industrial-scale recommendation systems, and serves as a backbone for Generative Recommenders.

\textbf{\emph{Adaptation to Pre-ranking.}} 
Since our method is applied in the pre-ranking stage of live streaming recommendation, while many baselines 
(including SASRec, ReaRec, and HSTU) were originally designed for the retrieval stage, a direct comparison 
would be inappropriate. To ensure fairness, we adapt these baselines to align with our setting. 
Specifically, when comparing with these baselines, we employ their backbone networks to encode user features, 
and similarly encode live room features. The resulting user and live room representations, along with cross 
features, are concatenated and fed into multiple MLPs for multi-task prediction. This adaptation ensures that 
the baselines and our method are evaluated under equivalent conditions.

\subsection{Implementation Details}
The static SIDs (\S~\ref{sec:static_sid}) and dynamic SIDs (\S~\ref{sec:dynamic_sid}) each consist of three levels, with 2,048 codebooks per level.  
The number of queries per task (\S~\ref{sec:sid_decoder}) is set to $2$.  
During inference, we use a beam size of $10$ (\S~\ref{sec:beam_fusion}).
The learning rate is set to $1\times10^{-3}$ and the weight decay to $1\times10^{-2}$.  
We train the model using the AdamW~\citep{loshchilov2017decoupled} optimizer.

\begin{figure}[t]
\centering
\includegraphics[width=0.85\columnwidth]{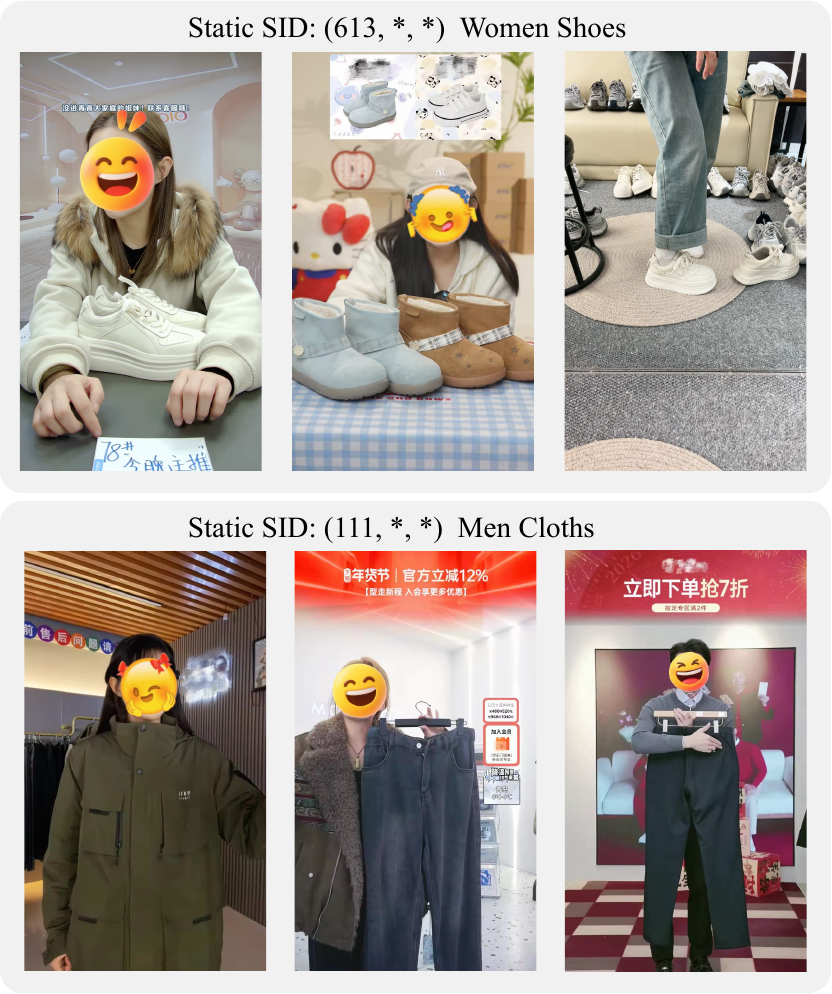}
\caption{
A case study of the static SID, where different prefixes correspond to live rooms in different categories. 
For example, (613, *, *) corresponds to the women's shoes category, and (111, *, *) corresponds to the men's clothing category.
}
\label{fig:static_sid_case}
\end{figure}

\section{Offline Experiments}
\label{appendix:offline_exp}
We provide additional offline analytical experiments.

\subsection{Case Study of Static SID}
\label{appendix:static_sid_case}
We conduct a case study to assess the effectiveness of static SIDs, as illustrated in Figure~\ref{fig:static_sid_case}. 
We examine the content of several live rooms associated with different static SID prefixes and observe that live rooms sharing the same prefix generally belong to the same category. 
This finding confirms that static SIDs effectively capture the categorical attributes of live streaming content.

\begin{table}[!t]
\centering
\caption{
Impact of the number of queries per task (\S~\ref{sec:sid_decoder}) on the overall performance.
}
\label{tab:offline_task_query_num}
\renewcommand{\arraystretch}{1.2}
\resizebox{1.0\columnwidth}{!}{
\begin{tabular}
{l
cccccc
}
\toprule
\multirow{2}{*}{\makecell[c]{\#Task \\ Query}} 
&\multicolumn{2}{c}{Watch30}
&\multicolumn{2}{c}{Watch200} 
&\multicolumn{2}{c}{Order} 
\\
\cmidrule(l){2-3}  \cmidrule(l){4-5} 
\cmidrule(l){6-7}  
&AUC &GAUC 
&AUC &GAUC 
&AUC &GAUC 
\\
\midrule
$Q=0$ & 0.7676 & 0.6929 & 0.8246 & 0.7266 & 0.8344 & 0.6960 \\
$Q=2$ & 0.7692 & 0.6956 & 0.8255 & 0.7281 & 0.8358 & 0.6946 \\
$Q=5$ & 0.7712 & 0.6976 & 0.8284 & 0.7315 & 0.8381 & 0.6956 \\
\bottomrule
\end{tabular}
} 
\end{table}

\subsection{Impact of Task Query Number}
\label{appendix:task_query}
We investigate the impact of the number of queries $Q$ per task in \S~\ref{sec:sid_decoder}, as shown in Table~\ref{tab:offline_task_query_num}. 
We compare three settings: no queries ($Q=0$), where the prediction is made directly from the decoder’s final hidden state, and $Q=2$ and $Q=5$.  
Results show that increasing the number of queries consistently improves performance, confirming the effectiveness of task queries in extracting task-relevant information to enhance overall accuracy.

\section{Online Experiments}

\subsection{Online Serving}
\label{appendix:online_serve}
In addition to the partial run optimization described in \S~\ref{sec:online_serve}, we employ the following engineering optimizations for online serving:

\textbf{Async Live Encoder.} 
As the live encoder is independent of user-specific information, we compute it asynchronously via external services. 
When the live room state changes, the encoder refreshes its output, which can be reused concurrently by tens of thousands of user requests.

\subsection{Efficiency Analysis}
\label{appendix:efficieny}
\begin{figure}[t]
\centering
\includegraphics[width=0.95\columnwidth]{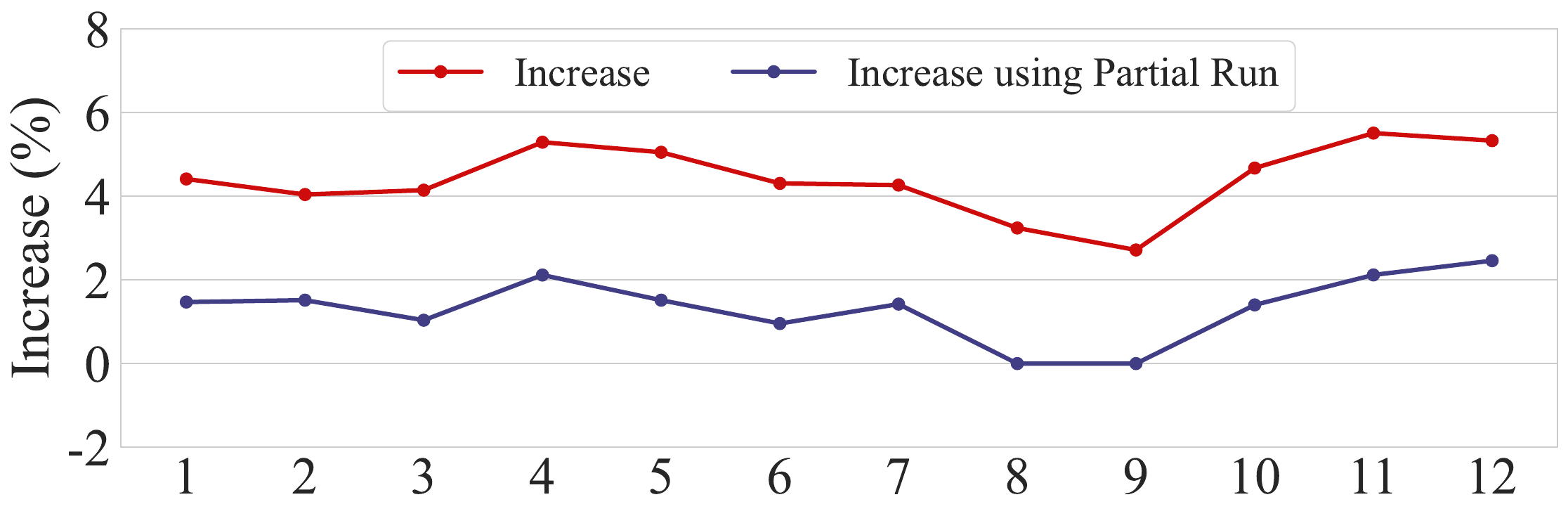}
\caption{
Latency increase of \ourname relative to DLRM in online serving at different time intervals, with and without the Partial Run optimization.
}
\label{fig:online_time}
\end{figure}

We further evaluate the end-to-end online inference latency of the recommendation pipeline for \ourname at different time intervals, as shown in Figure~\ref{fig:online_time}. 
Without engineering optimizations, latency increases by approximately $+4\%$. 
With the Partial Run optimization (\S~\ref{sec:online_serve}) enabled, the increase falls below $+2\%$, and in some cases reaches $0\%$, indicating no additional latency. 
Overall, the latency overhead is negligible, while \ourname consistently delivers substantial online performance gains.

\end{document}